\documentclass[preprint,prc,showpacs,preprintnumbers,amsmath,amssymb]{revtex4-1}
\usepackage{graphicx,latexsym,amssymb,amsmath,color,mathrsfs,booktabs,ifpdf}
\usepackage{epsfig,natbib}
\usepackage{amstext}
\usepackage{amsfonts}
\usepackage{dcolumn}
\usepackage{bm}
\usepackage{longtable}
\def\aA{$\alpha$-nucleus\ }
\def\aa{$\alpha+\alpha $\ }

\def\AA{nucleus-nucleus\ }
\def\aC{$\alpha+^{12}$C\ }
\def\Bea{$\alpha+^{8}$Be\ }
\def\BeA{$^{8}$Be-nucleus\ }
\def\BeBe{$^{8}$Be+$^{8}$Be\ }
\def\ac{$\alpha+^{12}$C }
\def\bebex{$^{8}$Be+$^{8}$Be$^*_{2^+}$\ }

\newcommand{\emax}{\mbox{$E_{\rm max}$}}
\newcommand{\lmax}{\mbox{$\ell_{\rm max}$}}

\newcommand{\alal}{\mbox{$\alpha+\alpha$}}
\begin{document}
\title{Coupled reaction channel study of the $^{12}$C($\alpha$,$^{8}$Be) reaction, 
 and the \BeBe optical potential}
\author{Do Cong Cuong$^1$}
\author{Pierre Descouvemont$^2$}
\author{Dao T. Khoa$^1$}
\author{Nguyen Hoang Phuc$^1$}
\affiliation{$^1$ Institute for Nuclear Science and Technology, 
 VINATOM, 100000 Hanoi, Vietnam. \\
$^2$ Physique Nucl\'eaire Th\'eorique et Physique Math\'ematique, C.P. 229,
	Universit\'e Libre de Bruxelles (ULB), B-1050 Brussels, Belgium}
\begin{abstract}
\begin{description}
\item[Background] Given the established 2$\alpha$ structure of $^{8}$Be, 
a realistic model of 4 interacting $\alpha$ clusters must be used to obtain
a \BeBe interaction potential. Such a four-body problem poses a challenge 
for the determination of the \BeBe optical potential (OP) that is still unknown 
due to the lack of the elastic \BeBe scattering data. 
\item[Purpose] To probe the complex \BeBe optical potential in the coupled reaction 
channel (CRC) study of the $\alpha$ transfer $^{12}$C$(\alpha,^{8}$Be) reaction 
measured at $E_\alpha=65$ MeV, and to obtain the spectroscopic information on 
the $\alpha+^{8}$Be cluster configuration of $^{12}$C.
\item[Method] The 3- and 4-body Continuum-Discretized Coupled Channel (CDCC) methods 
are used to calculate the elastic \Bea and \BeBe scattering at the energy around 
16 MeV/nucleon, with the breakup effect taken into account explicitly. Based on 
the elastic cross section predicted by the CDCC calculation, the local equivalent 
OP's for these systems are deduced for the CRC study of the $^{12}$C($\alpha,^{8}$Be) 
reaction.
\item[Results] Using the CDCC-based OP’s and $\alpha$ spectroscopic factors given 
by the cluster model calculation, a good CRC description of the $\alpha$ transfer 
data for both the \BeBe and  \bebex exit channels is obtained without any adjustment 
of the (complex) potential strength. 
\item[Conclusion] The \Bea and \BeBe interaction potential can be described 
by the 3- and 4-body CDCC methods, respectively, starting from a realistic \aa 
interaction. The $\alpha$ transfer $^{12}$C($\alpha,^{8}$Be) reaction should be 
further investigated not only to probe of the 4$\alpha$ interaction but also the 
cluster structure of $^{12}$C. 
\end{description}
\end{abstract}
\maketitle
\begin{center}
 Published in Phys. Rev. C {\bf 102}, 024622 (2020)
\end{center}

\section{Introduction}
The $\alpha$-cluster structure established for different excited states in several
light nuclei like $^{12}$C or $^{16}$O has inspired numerous experimental and 
theoretical studies, especially, the direct nuclear reactions measured with $^{12}$C 
as projectile and/or target \cite{Freer07}. Given the cluster states above the 
$\alpha$-decay threshold of $^{12}$C, some direct reactions with $^{12}$C were shown 
to produce both the free $\alpha$ particle and unstable $^{8}$Be in the exit channel 
\cite{Munoz10,Soy12,Curtis16}. Consequently, the knowledge of the \Bea and \BeBe 
interaction potentials should be important for the studies of such reactions within 
the distorted wave Born approximation (DWBA) or coupled reaction channel (CRC) 
formalism. 

Given a well established 2$\alpha$-cluster structure of the unbound $^{8}$Be nucleus, 
the \BeBe interaction potential poses a four-body problem which is a challenge for 
the determination of the \BeBe optical potential (OP) that cannot be deduced from 
a standard optical model (OM) analysis because of the lack of the elastic \BeBe 
scattering data. The knowledge about the \Bea and \BeA OP's should be also important 
for the studies of those direct reaction processes that produce $^8$Be fragments in 
the exit channel \cite{Rom09,PSM16,PSM19}. Although the \aA and \AA OP's are proven 
to be well described by the double-folding model (DFM) using the accurate ground-state 
densities of the colliding nuclei and a realistic density dependent nucleon-nucleon 
(NN) interaction (see, e.g., Refs.~\cite{Sat79,Kho97,Kho00,Kho01,Kho16}), the DFM 
cannot be used to calculate the \Bea and \BeBe OP's because of a strongly deformed,  
extended two-center density distribution of $^{8}$Be. In general, one could think 
of the triple- and quadruple folding models for the \Bea and \BeBe potentials, 
respectively, but these will surely be complicated and involve much more tedious 
calculation in comparison with the standard DFM method. Although some phenomenological 
OP's are available in the literature for $^7$Li and $^9$Be, two nuclei neighboring 
$^8$Be, a strongly (two-center) deformation of the unstable $^8$Be nucleus casts 
doubt on the extrapolated use of these potentials for the \BeBe and \BeA systems. 

Given a very loose (unbound) $^{8}$Be nucleus that breaks up promptly into 2 $\alpha$ 
particles, we determine in the present work the \BeBe OP using the Continuum-Discretized 
Coupled Channel (CDCC) method which was developed to take into account explicitly 
the breakup of the projectile and/or target. A textbook example is a direct reaction 
induced by deuteron, which is loosely bound and can be, therefore, easily broken up 
into a pair of free proton and neutron. Originally, the deuteron breakup states 
were included in terms of a discretized continuum by the CDCC method (see, e.g., 
Refs.~\cite{KYI86,AIK87,YOM12} for reviews). 
In the recent version of the CDCC theory, the continuum of deuteron is approximated 
by the square-integrable functions corresponding to positive energies. As a result, 
this approach can be well extended to study elastic scattering of exotic nuclei 
that have rather low breakup energies (typical examples are $^6$He and $^{11}$Be).

The first developments of the CDCC method were done in the framework of a three-body 
system where the projectile is seen as a two-body nucleus and target is assumed 
to be structureless, being in its ground state. More recently, four-body calculations 
were developed, for either a three-body projectile on a structureless target \cite{MHO04}, 
or a two-body projectile and a two-body target \cite{De18}. The latter approach 
is highly time consuming, but was successfully applied to study $^{11}{\rm Be}+d$ 
scattering in terms of $^{11}{\rm Be}=^{10}$Be$+n$ and $d=p+n$. The goal of the present 
study is to determine the \BeBe OP based on the elastic scattering matrix predicted
by the 4-body CDCC calculation of four interacting $\alpha$ clusters. While such a 
$4\alpha$ model is not appropriate for the spectroscopy of $^{16}$O \cite{De18}, 
the derived OP for elastic \BeBe scattering is expected to be reliable. The only input 
for the present $4\alpha$ CDCC calculation is a realistic \aa interaction potential.

Although $^8$Be is particle-unstable, its half life around $10^{-16}$s is long enough 
for the $^8$Be-nucleus OP to contribute significantly to a direct reaction that 
produces $^8$Be in the exit channel, like the $\alpha$ transfer 
$^{12}$C($\alpha$,$^8$Be) reaction. This particular reaction was shown to be a good 
tool for the study of the high-lying or resonance states of $^{16}$O \cite{Che67,Mar72} 
and to determine the $\alpha$ cluster configurations of this nucleus 
\cite{Che67,Bro76,Cur13}. Because of the unbound structure of $^8$Be, the direct 
reaction reactions $A(\alpha$,$^8$Be)$B$ usually have a very low cross section (of a 
few tens microbarn), but they are extremely helpful for the study of the $\alpha$-cluster 
structure of the target nuclei \cite{Woz76}. In particular, the $\alpha$ spectroscopic 
factors of different cluster states were deduced from these measurements at the 
$\alpha$ incident energies of 65 to 72.5 MeV. 

In the present work, the \Bea and \BeBe optical potentials deduced from the scattering 
wave functions given by the 3- and $4\alpha$ CDCC calculations are used as the core-core 
and the exit OP, respectively, in the CRC study of $^{12}$C($\alpha$,$^8$Be) reaction 
measured at 65 MeV \cite{Woz76}. The OP of the entrance channel is calculated in the 
DFM using the density dependent CDM3Y6 interaction that was well tested in the 
mean-field studies of nuclear matter as well as in the OM studies of the elastic \aA 
scattering \cite{Kho97,Kho01}, and it accounts well for the elastic \aC scattering 
data measured at 65 MeV \cite{Yasue83,Gon14}. The $\alpha$ spectroscopic factors 
of the $\alpha+^{8}$Be cluster configurations of $^{12}$C are taken from the results 
of the complex scaling method (CSM) by Kurokawa and Kato \cite{KaKu07}.

\section{Three- and four-body CDCC methods}
\label{sec2}
We discuss here the CDCC method used to determine the \Bea and \BeBe optical 
potentials, where the $\alpha$ particles are treated as structureless and interacting 
with each other through a (real) potential $v_{\alpha\alpha}(r)$. The Hamiltonian 
of the $\alal$ system is given by
\begin{equation}
 H_{\alpha\alpha}(\bm{r})=T_r+v_{\alpha\alpha}(r),
\label{eq1}
\end{equation}
where $T_r$ is the relative kinetic energy. There are two versions of the \aa potential
\cite{AB66,BFW77} parametrized in terms of Gaussians amenable for the present CDCC 
calculation. In the present work, we have chosen the \aa potential suggested by Ali 
and Bodmer \cite{AB66} (referred to hereafter as AB potential). The AB potential 
simulates the Pauli blocking effect by a repulsive core that makes this potential 
much shallower than the deep \aa potential suggested by Buck {\it et al.} \cite{BFW77}. 
Both potentials reproduce equally well the \aa phase shifts, and they were shown 
by Baye \cite{Ba87} to be linked by a supersymmetric transformation. The Buck potential, 
however, contains some deeply-bound states (two states with $\ell=0$ and one with 
$\ell=2$), which do not have physical meaning but simulate the so-called Pauli forbidden 
states \cite{Sa69} in the microscopic $\alpha+\alpha$ model. As long as the two-body 
$\alpha+\alpha$ system is considered, the choice of either AB or Buck potential is not 
crucial. However, when dealing with more than two $\alpha$ clusters, the forbidden 
states have to be removed as they produce spurious states in a multi-cluster system 
like \Bea or $^{8}$Be+$^{8}$Be. There are two methods to remove the forbidden states 
in the multi-cluster systems: either to apply the pseudostate  method \cite{Ne71} or 
to use the supersymmetric transformation \cite{Ba87}. These two techniques, however, 
give rise to a strong angular-momentum dependence of the \aa potential that cannot be used 
in most of the multi-cluster models. Among the 3$\alpha$ models, only the hypersperical 
method and Faddeev method are able to use the deep \aa potential with an exact 
removal of the \aa forbidden states. This is why other studies of the 3$\alpha$ 
and 4$\alpha$ systems \cite{Og09,De18,Su02,Ti08} have used only the $\ell$-independent 
AB potential. In the present work, we perform the CDCC calculation of the \Bea and 
\BeBe optical potentials using the AB potential of the \aa interaction, so that 
the spurious effects arising from the Pauli forbidden states can be avoided.

The present CDCC method is based on the eigenstates $\Phi_{\lambda}^{\ell m}(\bm{r})$ 
of the Hamiltonian (\ref{eq1}) which can be written as
\begin{align}
\Phi_{\lambda}^{\ell m}({\bm r})=\frac{1}{r}{u_{\lambda}^{\ell}(r)}
 Y_{\ell m}(\hat{\bm r}), \nonumber
\end{align}
where $\ell$ is the relative orbital momentum of the \aa system. The radial wave 
functions $u_{\lambda}^{\ell}(r)$ of the two-$\alpha$ state $\lambda$ are expanded over 
a basis of $N$ orthonormal functions $\varphi_i(r)$ 
\begin{align}
 u_{\lambda}^{\ell}(r)=\sum_{i=1}^N f_{\lambda,i}^{\ell}\varphi_i(r),
\label{eq2}
\end{align}
where $f_{\lambda,i}^{\ell}$ are determined by diagonalizing the eigenvalue problem 
\begin{align}
\sum_j f_{\lambda,j}^{\ell}\bigl(\langle\varphi_i|H_{\alpha\alpha}|\varphi_j\rangle 
 - E_{\lambda}^{\ell}\delta_{ij}\bigr)=0. \label{eq3}
\end{align}
The eigenvalues with $E_{\lambda}^{\ell}<0$ correspond to the physical bound states, 
while those with $E_{\lambda}^{\ell}>0$ are referred to as the pseudostates, 
which are used in the present CDCC method to simulate the breakup of $^8$Be. 
Note that there is only a small number of physical states in a CDCC calculation 
(often one for exotic nuclei). Although $^8$Be is unbound, its energy is very 
close to the \aa threshold, and the lifetime is long enough to use a quasi-bound
approximation for the ground state (g.s.). Equations (\ref{eq2}) and (\ref{eq3}) 
are general for any choice of the basis functions $\varphi_i(r)$. We use 
here a Lagrange-mesh basis \cite{Ba15} derived from the Legendre polynomials, and
the calculation of matrix elements in Eq.~(\ref{eq3}) is fast and accurate. We refer 
the reader to Ref.~\cite{Ba15} for more details and the application of the 
Lagrange-mesh basis. 

\begin{figure}[htb]
	\begin{center}
		\epsfig{file=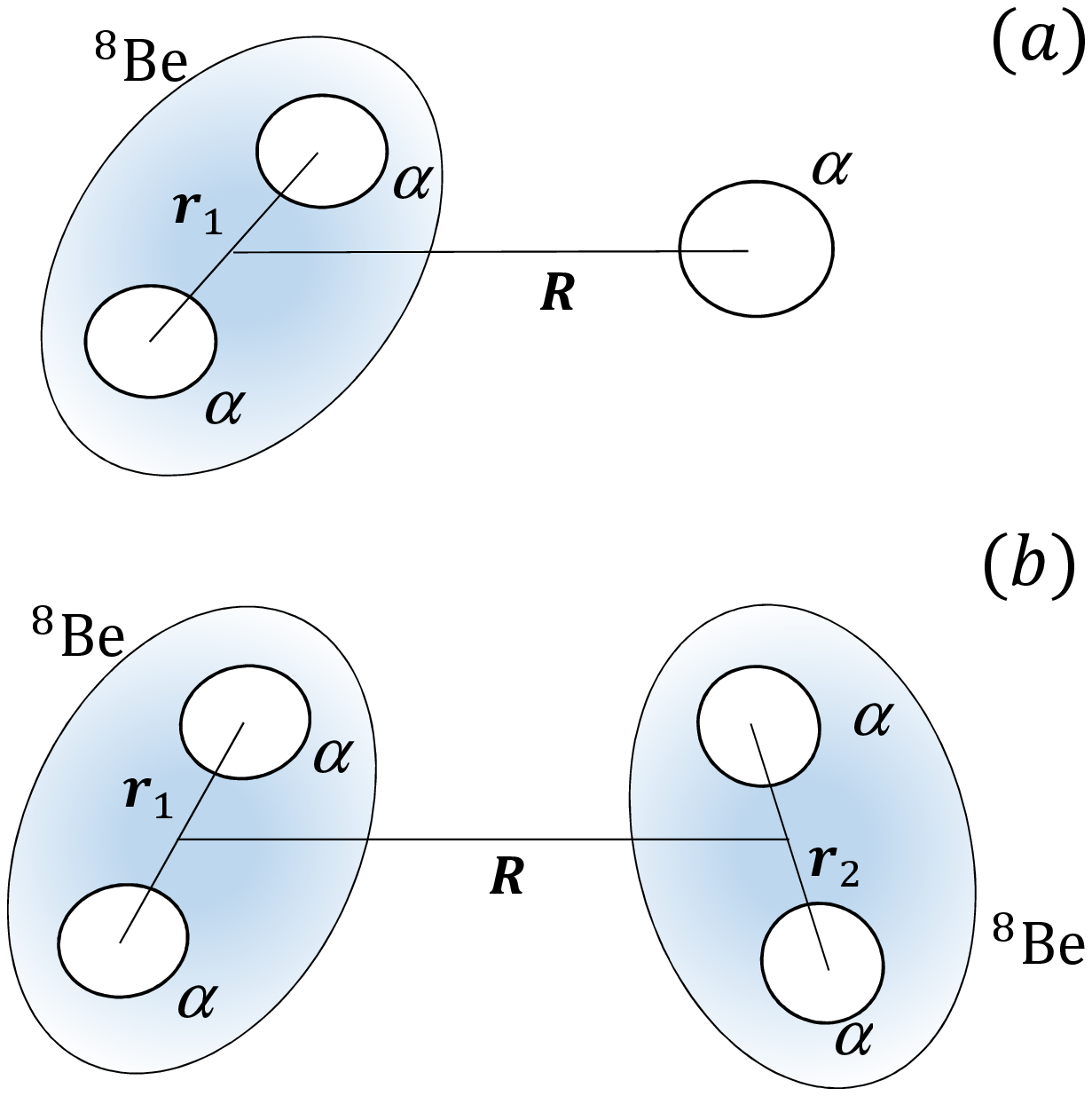,width=7.5cm}
		\caption{Alpha configurations and coordinates used for \Bea (a) and for \BeBe (b).}
		\label{f1}
	\end{center}
\end{figure}

The \Bea and \BeBe systems are described by the 3$\alpha$ and 4$\alpha$ Hamiltonians, 
respectively, as
\begin{align}
& H_3=H_{\alpha\alpha}(\pmb{r}_1)+T_R+\sum_{i=1}^2 v_{\alpha\alpha}(S_{1i}) \nonumber \\
& H_4=H_{\alpha\alpha}(\pmb{r}_1)+H_{\alpha\alpha}(\pmb{r}_2)+
T_R+\sum_{i,j=1}^2 v_{\alpha\alpha}(\vert \pmb{S}_{1i}-\pmb{S}_{2j}\vert), 
\label{eq4}
\end{align}
where $\bm{R}$ is the projectile-target coordinate and $(\bm{r}_1,\bm{r}_2)$ are 
the internal coordinates of the $^8$Be nuclei, as illustrated in Fig.~\ref{f1}. 
Coordinates $\pmb{S}_{1i}$ and $\pmb{S}_{2j}$ are expressed as a function 
of $(\pmb{R},\pmb{r}_1)$ and of $(\pmb{R},\pmb{r}_2)$, respectively.

In the CDCC approximation, the \Bea wave function for a total angular moment $J$ 
and parity $\pi$ is written as
\begin{align}
\Psi^{JM\pi}(\bm{R},\bm{r}_1)=\sum_{\lambda\ell L}g^{J\pi}_{\lambda \ell L}(R)
\bigl[\Phi_{\lambda}^{\ell}(\bm{r}_1)\otimes Y_L(\hat{\bm R})\bigr]_{JM}, \label{eq5}
\end{align}
where the summation over the pseudostates $\lambda$ is truncated at a given energy 
$\emax$. The \BeBe wave functions involve 4 $\alpha$ clusters and can be expressed as
\begin{align}
&\Psi^{JM\pi}(\bm{R},\bm{r}_1,\bm{r}_2)=\sum_{\lambda_1 \ell_1}\sum_{\lambda_2\ell_2}
 \sum_{IL}g^{J\pi}_{\lambda_1 \ell_1 \lambda_2 \ell_2 I L}(R)\nonumber \\
&\hspace*{1cm} \times \biggl[\bigl[\Phi_{\lambda_1}^{\ell_1}(\bm{r}_1) 
\otimes\Phi_{\lambda_2}^{\ell_2}(\bm{r}_2)\bigr]_I 
\otimes Y_L(\hat{\bm R}) \biggr]_{JM},
\label{eq6}
\end{align}
where the parity conservation imposes $(-1)^L=(-1)^I$. There are two parameters defining 
the CDCC basis: the maximum $^8$Be angular momentum $\lmax$ and maximum pseudostate 
energy $\emax$. The physical quantities obtained from the CDCC calculation (scattering 
matrix, elastic cross section, and local equivalent OP) must be converged with respect 
to these parameters. In practice, the 4$\alpha$ calculations involve many channels and 
are, therefore, highly time consuming \cite{De18}.

The relative radial wave functions $\chi^{J\pi}_c(R)$, with the indices 
$c=(\lambda\ell L)$ for \Bea and $c=(\lambda_1 \ell_1 \lambda_2 \ell_2 I L)$ for \BeBe, 
are obtained from the solutions of the following coupled-channel equations
\begin{align}
&\biggl[-\frac{\hbar^2}{2\mu}\biggl(\frac{d^2}{dR^2}-\frac{L(L+1)}{R^2}  \biggr)
+E_{c_1}+E_{c_2}-E_{\rm c.m.} \biggr]\chi^{J\pi}_{c}(R)\nonumber \\
&\hspace*{1cm} +\sum_{c'}V^{J\pi}_{cc'}(R)\chi^{J\pi}_{c'}(R)=0,
\label{eq7}
\end{align}
where $\mu$ is the reduced mass, $E_{\rm c.m.}$ is the center-of-mass energy, 
$E_{c_1}$ and $E_{c_2}$ are the excitation energies of the two interacting nuclei, 
separated by the distance $R$ as shown in Fig.~\ref{f1}. The coupling potentials 
$V^{J\pi}_{cc'}(R)$ are determined by the method explained in Refs.~\cite{RAG08,De18}. 
The system of the coupled channel equations (\ref{eq7}) is solved using the $R$-matrix 
method which provides explicitly the scattering matrix and the associated wave 
functions \cite{DB10,De16a}. Although the AB potential \cite{AB66} is real, it 
consistently reproduces the experimental \aa phase shifts up to about 20 MeV. 
In the present CDCC approach, the loss of flux from the elastic scattering channel 
is due entirely to the breakup channels, and the local equivalent \Bea and \BeBe 
optical potentials are therefore complex. Owing to the strong 2$\alpha$ structure 
of $^8$Be it is likely that these breakup channels represent the main source 
of the absorption.

\section{Results and discussion}
\label{sec3}
\subsection{Local equivalent OP for the \Bea and \BeBe systems}
\label{sec3.1}
The main goal of our study is to determine the local ($J$-independent) equivalent 
optical potential $U$ for the \Bea and \BeBe systems at the considered energies, 
based on the scattering wave functions given by the solutions of the CDCC equations 
(\ref{eq7}). The main requirement for this procedure is that the solutions 
$\chi^{J\pi}$ of the one-channel OM equation with the optical potential $U$ 
\begin{align}
\bigl[E_{\rm c.m.}-T_R-U(R)\bigr]\chi^{J\pi}(R)=0 \label{eq8}
\end{align}
give the cross section of the elastic \Bea or \BeBe scattering close to that 
given by the 3$\alpha$ or 4$\alpha$ CDCC calculation (\ref{eq7}), especially, 
the cross section at forward angles which is sensitive to the surface part 
of the 3$\alpha$ or 4$\alpha$ interaction potential. We briefly discuss the 
two approaches used in the present work for this purpose. 
\begin{figure}[h]\vspace*{-1cm}
	\includegraphics[width=1.0\textwidth]{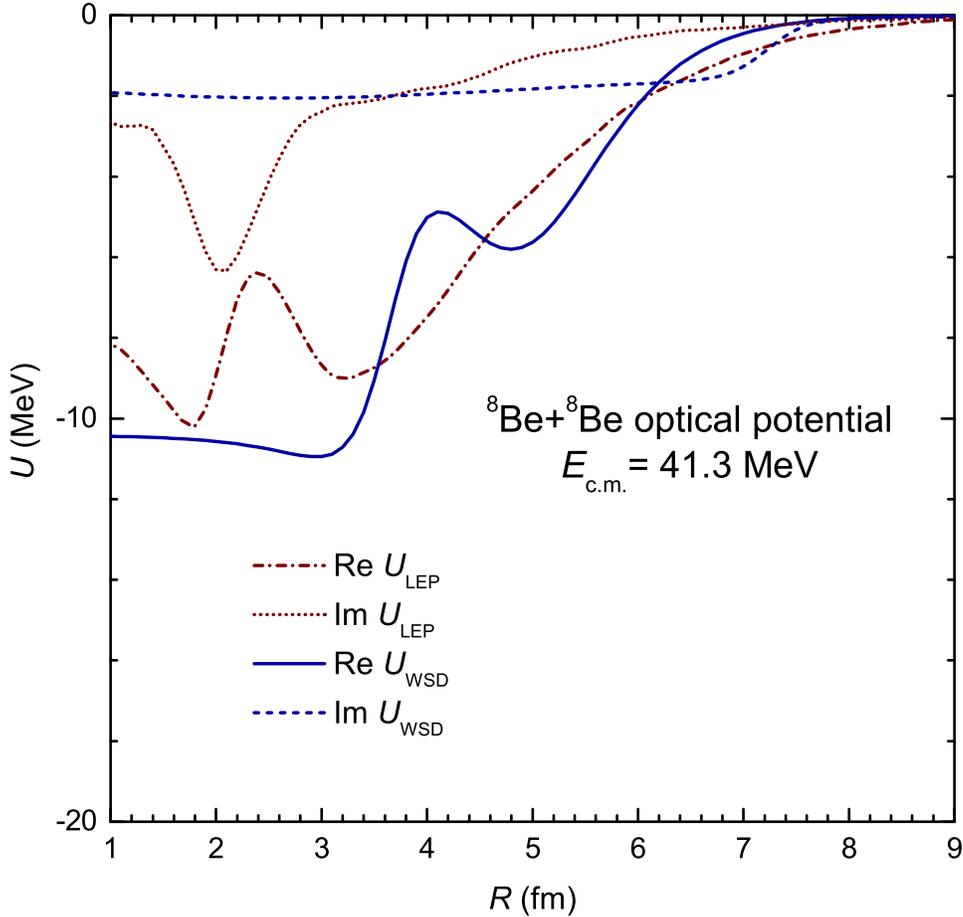}\vspace*{-1cm} 
\caption{Complex OP for the elastic \BeBe scattering at $E_{\rm c.m.}=41.3$ MeV 
given by the method (i), and that in the WS form (\ref{eq9}) given by the method 
(ii).}  \label{f2}
\end{figure}

(i) The quantum-mechanically consistent method using the matrix inversion was suggested 
in Refs.~\cite{TNL89,De18} to derive a local equivalent potential (LEP) that exactly 
reproduces the elastic cross section given by the CDCC calculation (\ref{eq7}). 
However, this LEP has two major drawbacks that prevent its further use in the direct 
nuclear reaction calculation. Namely, the derived LEP strongly depends on the total 
angular momentum $J$, and its radial dependence has the singularities caused by the 
nodes of the scattering wave functions. These problems can be handled by the method 
proposed in Ref.~\cite{TNL89} which averages the obtained LEP over the angular momenta 
to obtain a smooth $J$-independent OP without discontinuity that approximately 
reproduces the CDCC elastic scattering cross section. The recent 4-body CDCC 
calculation \cite{De18} has shown that such an averaging method to a fairly 
good approximation determines the local $J$-independent OP. The complex OP 
derived using this approach is denoted hereafter as $U_{\rm LEP}$, with its imaginary 
part $W_{\rm LEP}$ originating from the breakup channels included in the CDCC 
calculation (\ref{eq7}). We have first performed the CDCC calculation (\ref{eq7}) 
for the elastic \Bea and \BeBe scattering at $E_{\rm c.m.}=43.3$ and 41.3 MeV, 
respectively, using the AB potential \cite{AB66} for the \aa interaction. 
The maximum angular momentum of the \aa system is $\lmax=2$, 
and the maximum pseudostate energy is $\emax=10$ MeV. These cutoff values were 
well tested to ensure the convergence of both the elastic cross section and 
$U_{\rm LEP}$. The complex $U_{\rm LEP}$ for the \Bea and \BeBe systems were 
obtained first in a Lagrange mesh \cite{TNL89}, and then interpolated into the smooth 
shapes for use as the external input of the complex OP in the CRC calculation. 
\begin{figure}[h]\vspace*{0cm}
 \includegraphics[width=0.75\textwidth]{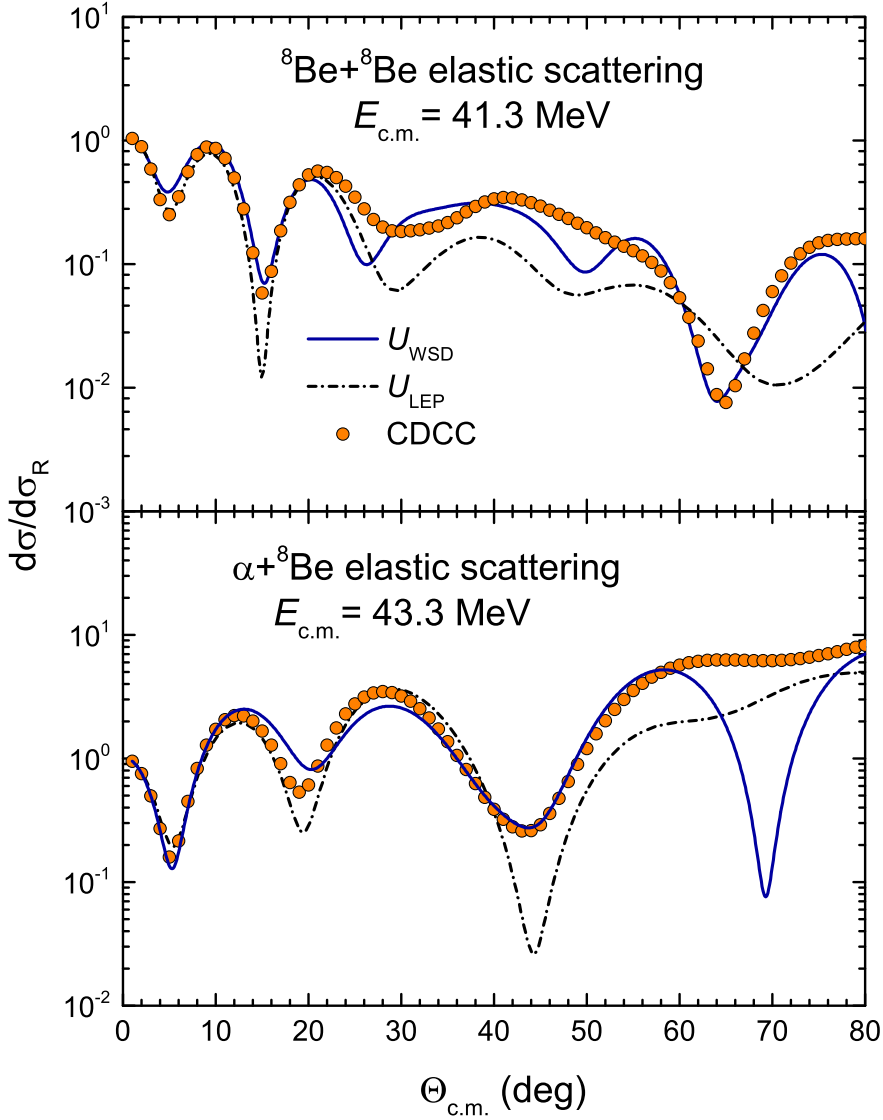}\vspace*{-0.5cm} 
 \caption{The CDCC prediction for the elastic \BeBe (upper panel) and \Bea (lower 
 panel) scattering at $E_{\rm c.m.}=41.3$ and 43.3 MeV, in comparison  with the 
 corresponding OM results given by the optical potentials $U_{\rm LEP}$  and 
 $U_{\rm WSD}$ determined by the methods (i) and (ii), respectively.}	\label{f3}
\end{figure}

(ii) The standard OM method can also be used to determine a phenomenological 
$J$-independent OP in the conventional Woods-Saxon (WS) form, with its parameters 
adjusted to obtain a good OM fit to the elastic cross section given by the 3$\alpha$ 
or 4$\alpha$ CDCC calculation. We have assumed in the present work the following 
(volume+surface) WS form for the complex OP of the \Bea and \BeBe systems at the 
energies under study, denoted hereafter as $U_{\rm WSD}$ 
\begin{align}
- & U_{\rm WSD}(R)=V_vf_v(R)-4V_da_d\frac{df_d(R)}{dR}+i\left[W_vf_w(R)
 -4W_da_s\frac{df_s(R)}{dR}\right], \nonumber\\
 & {\rm where}\ f_x(R)=\frac{1}{1+\exp[(R-R_x)/a_x]},\ x=v,d,w,s.  
\label{eq9}
\end{align}
The elastic \BeBe and \Bea cross sections at $E_{\rm c.m.}=41.3$ and 43.3 MeV, 
predicted by the 4$\alpha$ and 3$\alpha$ CDCC calculations (\ref{eq7}), respectively,
have been used in the method (ii) as the ``experimental data" with the uniform 10\% 
uncertainties for the OM analysis to determine the WS parameters of $U_{\rm WSD}$. 
We found quite a shallow WS potential that gives a good agreement of the OM result
with the CDCC elastic cross section (see the OP parameters in Table~\ref{t1}).
\begin{table}[h]
\setlength{\tabcolsep}{0.3em}
\centering
\caption{WS parameters (\ref{eq9}) of $U_{\rm WSD}$ given by the method (ii) based 
 on the OM fit to the elastic \BeBe and \Bea cross sections at $E_{\rm c.m.}=41.3$ 
and 43.3 MeV, predicted by the 4$\alpha$ and 3$\alpha$ CDCC calculations 
(\ref{eq7}), respectively. $J_V$ and $J_W$ are the volume integrals per interacting 
nucleon pair of Re~$U_{\rm WSD}$ and Im~$U_{\rm WSD}$, respectively. $J_{V_{\rm LEP}}$ 
and $J_{W_{\rm LEP}}$ are those of $U_{\rm LEP}$ given by the method (i).}
\label{t1}\vspace{0.5cm}
\begin{tabular}{ccccccccc} 
 \hline\hline
 & $V_v(W_v)$ & $R_{v(w)}$ & $a_{v(w)}$ & $V_d(W_d)$ & $R_{d(s)}$ & $a_{d(s)}$ & 
 $-J_V(J_W)$ & $-J_{V_{\rm LEP}}(J_{W_{\rm LEP}})$\\ 
 & (MeV) & (fm) & (fm) & (MeV) & (fm) & (fm) & (MeV~fm$^3$) & (MeV~fm$^3$) \\ \hline 
 \multicolumn{9}{c}{$^8$Be+$^8$Be, $E_{\rm c.m.}=41.3$ MeV} \\ \hline
  Real & 10.42 & 3.652 & 0.190 & 5.785  & 4.811  & 0.562 & 95.88 & 100.0 \\             
  Imag. & 1.486 & 7.225 & 0.183 & 0.567  & 2.665  & 1.564 & 46.87 & 29.71 \\ \hline
    \multicolumn{9}{c}{$^4$He+$^8$Be, $E_{\rm c.m.}=43.3$ MeV}  \\ \hline
  Real  & 1.535 & 5.182 & 0.123 & 5.397 & 2.584 & 0.997 & 111.0 & 97.62 \\   
  Imag. & 10.09 & 1.771 & 0.125 & 8.201 & 3.299 & 0.120 & 24.56 & 33.95 \\ 
	\hline\hline
	\end{tabular}
\end{table}

The radial shapes of both $U_{\rm LEP}$ and $U_{\rm WSD}$ potentials for the \BeBe 
system at $E_{\rm c.m.}=41.3$ MeV are shown in Fig.~\ref{f2}, the use of the shallow 
AB potential of the \aa interaction is shown to result on quite a shallow potential 
$U_{\rm LEP}$. A moderate oscillation of $U_{\rm LEP}$ is seen at small radii that 
might originate from the $J$-dependence of the exact LEP discussed above. The best-fit 
WS complex OP determined by the method (ii) has the strength of Re~$U_{\rm WSD}$ 
enhanced slightly at the surface ($R\approx 5$ fm), and a weak and smooth 
Im~$U_{\rm WSD}$. One can see in Table~\ref{t1} that the volume integrals of 
Re~$U_{\rm WSD}$ and Im~$U_{\rm WSD}$ are close to those of Re~$U_{\rm LEP}$ and 
Im~$U_{\rm LEP}$, which indicates that the OP's given by both methods belong to about 
the same potential family. The results of the CDCC calculation (\ref{eq7}) for 
the elastic \BeBe and \Bea scattering at $E_{\rm c.m.}=41.3$ and 43.3 MeV, respectively, 
are compared in Fig.~\ref{f3} with the results of the OM calculation (\ref{eq8}) 
using $U_{\rm LEP}$ and $U_{\rm WSD}$. The OM results given by both OP's agree 
fairly good with the CDCC prediction at forward angles, while at medium and large angles
the phenomenological $U_{\rm WSD}$ determined by the method (ii) better reproduces 
the CDCC cross sections. The agreement with the CDCC results becomes worse at large 
angles, and it might be due to the nonlocality effects. 

We note further that the method (i) fails to derive a smooth $\ell$-independent 
$U_{\rm LEP}$ based on the CDCC results obtained with the deep Buck \aa potential. 
Namely, the obtained $\ell$-independent $U_{\rm LEP}$ turns out to be deeper but 
strongly oscillatory, and it gives the elastic cross section substantially different 
from that given by the CDCC calculation. Such a failure of the $\ell$-independent 
$U_{\rm LEP}$ based on the Buck potential is presumably caused by the Pauli forbidden 
states, and this remains an unsolved problem for the present 4$\alpha$ CDCC method. 
Therefore, we deem hereafter reliable only the CRC results obtained with \Bea and \BeBe 
OP's derived based on the CDCC elastic cross section obtained with the AB potential
of the \aa interaction. 
     
\subsection{CRC study of the $^{12}$C($\alpha,^8$Be) reaction}
\label{sec3.2}
The \Bea and \BeBe optical potentials determined by the methods (i) and (ii) have been 
further used as the potential inputs for the CRC study of the $\alpha$ transfer 
$^{12}$C($\alpha$,$^8$Be) reaction measured at $E_\alpha=65$ MeV \cite{Woz76}. 
We briefly recall the multichannel CRC formalism, to illustrate how the OP's of the  
\Bea and \BeBe systems enter the CRC calculation of the $\alpha$ transfer cross section. 
In general, the CRC equation for the initial channel $\beta$ of the transfer reaction 
can be written as \cite{Sat83,ThompsonN}
\begin{eqnarray}
&&\hskip 1cm \bigl[E_{\beta}-T_{\beta}-U_{\beta}({\bm R})\bigr]
 \chi_{\beta}({\bm R})= \nonumber\\
&&\sum_{\beta'\neq\beta}
 \bigl\{\langle\beta|W|\beta'\rangle +\langle\beta|\beta'\rangle
 \bigl[T_{\beta'}+U_{\beta'}({\bm R}')-E_{\beta'}\bigr]\bigr\}
 \chi_{\beta'}({\bm R}'). \label{eq10}
\end{eqnarray}
\begin{figure}[bht]\vspace*{-2cm}
 \hspace*{-1cm}\includegraphics[width=0.8\textwidth]{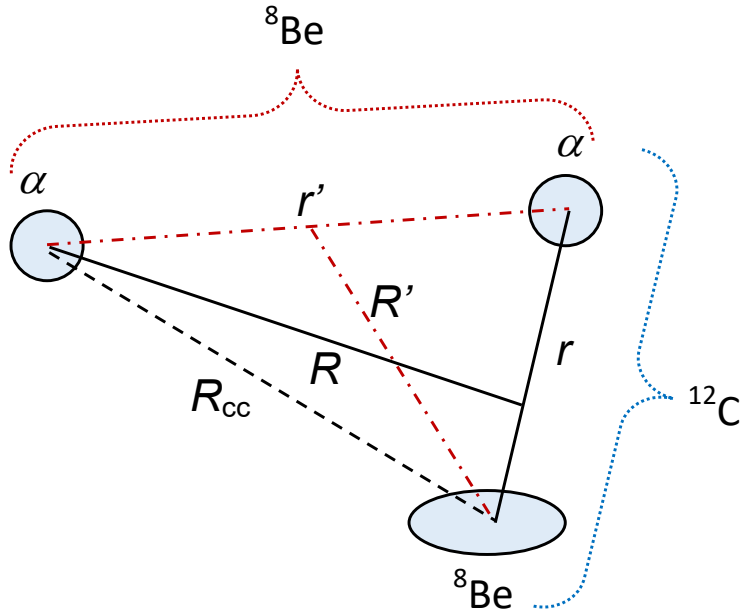}\vspace*{-2.0cm} 
 \caption{Cluster configurations in the entrance and exit channels of the 
 $^{12}$C($\alpha,^8$Be) reaction, and the corresponding coordinates used 
 for the inputs of the potentials in the CRC calculation.} 
\label{f4}
\end{figure}
Without coupling to the inelatic scattering channels, the indices $\beta$ and $\beta'$ 
in Eq.~(\ref{eq10}) stand for the initial \aC and final \BeBe partitions of the $\alpha$ 
transfer reaction, respectively, as shown in Fig.~\ref{f4}. In the present CRC analysis, 
the index $\beta'$ is used to identify both the $^{8}$Be+$^{8}$Be$_{\rm g.s.}$ and 
$^{8}$Be+$^{8}$Be$^*_{2^+}$ exit channels of the final partition. The distorted waves 
$\chi_{\beta}$ and $\chi_{\beta'}$ are given by the optical potentials $U_{\beta}$ 
and $U_{\beta'}$ of the \aC and \BeBe systems, respectively. The $\alpha$ transfer 
proceeds via the transfer interaction $W$ which is determined in the post form 
\cite{Sat83,ThompsonN} as  
\begin{equation}
 W=V_{\alpha - ^{12}\text{C}}(\bm{r})+\bigl[U_{\alpha+^8\text{Be}}(\bm{R}_{\rm cc})
 -U_{^8\text{Be}+^8\text{Be}}(\bm{R}')\bigr], \label{eq11}
\end{equation}
with the radii of the potentials illustrated in Fig.~\ref{f4}. Here 
$V_{\alpha-^{12}\text{C}}(r)$ is the potential binding the $\alpha$ cluster 
to the $^8$Be core in the g.s. of $^{12}$C. The difference between the core-core OP 
and that of the final partition, 
$U_{\alpha+^8\text{Be}}(R_{\rm cc})-U_{^8\text{Be}+^8\text{Be}}(R')$, 
is the complex remnant term of $W$. The CRC equations (\ref{eq10}) are solved 
iteratively using the code FRESCO written by Thompson \cite{Thompson}, with the 
complex (nonlocal) remnant term and boson symmetry of the identical \BeBe
system properly taken into account. One can see that the \BeBe OP enters the CRC 
calculation of the $\alpha$ transfer $^{12}$C($\alpha$,$^8$Be) reaction as the input 
of both the remnant term and the OP of the final partition. Therefore, it can be 
tested indirectly based on the CRC description of the $\alpha$ transfer data. 

\begin{figure}[bht]\vspace*{-0.5cm}
 \includegraphics[width=\textwidth]{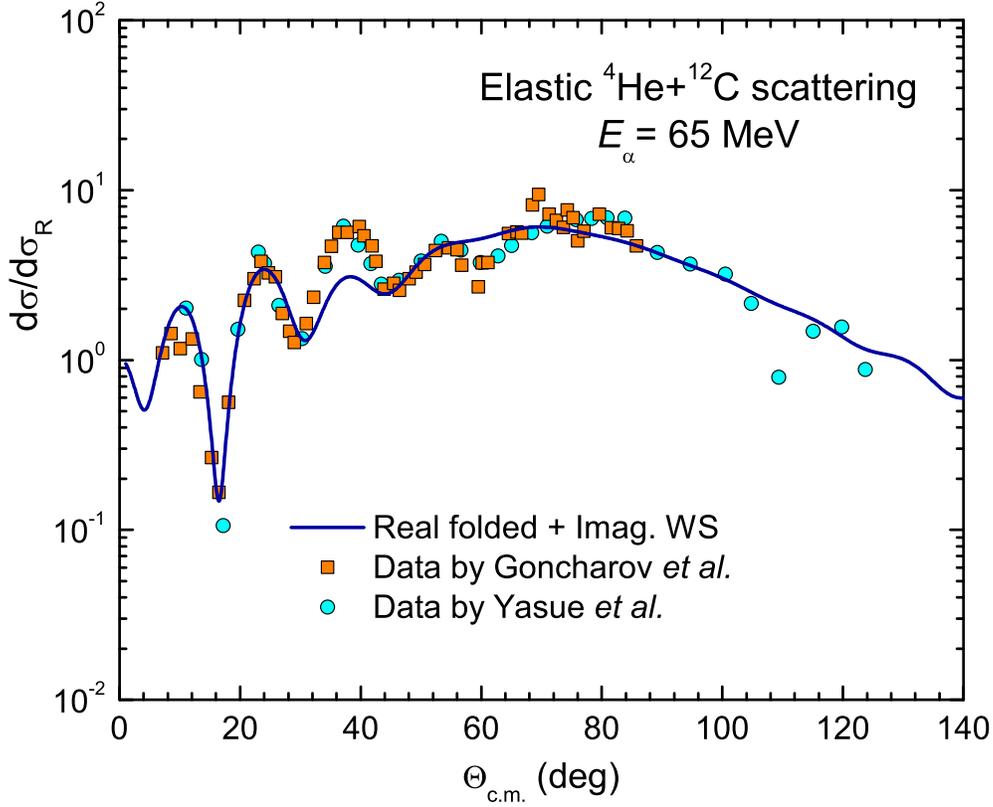}\vspace*{-0.5cm} 
 \caption{CRC description of the elastic \ac scattering at $E_{\alpha}=65$ 
 MeV given by the (unrenormalized) real folded OP and imaginary OP chosen
 in the WS form \cite{Kho01}, in comparison with the data taken from 
 Refs.~\cite{Yasue83,Gon14}} \label{f5}
\end{figure}
The OP of the initial partition $U_{\alpha+^{12}\text{C}}$ has its real part given 
by the double-folding model using the density dependent CDM3Y6 interaction \cite{Kho01}, 
and imaginary part chosen in the WS shape, with the parameters fine tuned to the 
best CRC fit of the elastic \aC scattering data measured at $E_\alpha=65$ MeV 
\cite{Yasue83,Gon14}. A reasonably good CRC description of the elastic \ac scattering 
data at $E_{\alpha}=65$ MeV (see Fig.~\ref{f5}) is achieved without renormalizing
the strength of the real OP. In principle, we could also think of using the
4$\alpha$ CDCC method to predict the \ac OP at the considered energy. However, 
Suzuki {\it et al.} \cite{Suzuki08} have shown that the use of a \emph{local} \aa potential 
(that properly reproduces the experimental \aa phase shifts) cannot provide a proper 
3$\alpha$ description of both the g.s. and $0^+_2$ excitation (known as Hoyle state) 
of $^{12}$C. This problem could only be solved by introducing a microscopically 
founded \emph{nonlocal} \aa force that mimics the interchange of three $\alpha$ clusters 
in the phase space allowed by the Pauli principle \cite{Suzuki08}. The use a nonlocal 
\aa interaction remains beyond the scope of the present 4-body CDCC method \cite{De18}.
On the other hand, the elastic \ac scattering at energies above 10 MeV/nucleon 
is proven to be strongly refractive \cite{Kho97,Kho01,Kho07r}, with a far-side 
dominant elastic cross section at large angles typical for the nuclear rainbow, 
which can be well described by the deep (mean-field type) real OP predicted 
by the double-folding model \cite{Kho01,Kho07r}.  

For the $\alpha$ transfer reaction, the initial (internal) state of the $\alpha$ 
cluster bound in $^{12}$C is assumed to be $1s$ state. Then, the relative-motion wave 
function $\Phi_{NL}(\bm{r})$ of the $\alpha$+$^{8}$Be configuration in the $^{12}$C 
target ($L$-wave state) has the number of radial nodes $N$ determined by the Wildermuth 
condition \cite{Sat83}, so that the total number of the oscillator quanta 
$\mathcal{N}$ is conserved  
\begin{equation}
 \mathcal{N}=2(N-1)+L=\sum_{i=1}^{4}2(n_i-1)+l_i, \label{eq12}
\end{equation}
where $n_i$ and $l_i$ are the principal quantum number and orbital momentum of each 
constituent nucleon in the $\alpha$ cluster. $\Phi_{NL}(\bm{r})$ is obtained in the
potential model using $V_{\alpha - ^{12}\text{C}}(r)$ chosen in the WS shape, with its 
radius and diffuseness fixed as $R=3.767$ fm and $a=0.65$ fm, and the WS depth 
($V=51.6$ MeV) adjusted to reproduce the $\alpha$ separation energy of $^{12}$C. 
Because the ground state of $^8$Be is unbound by 92 keV, we have used in the present 
CRC calculation a {\it quasi-bound} approximation for $^8$Be similar to that used for 
the g.s. of $^8$Be in the CDCC calculation as discussed in Sec.~\ref{sec2}, to 
describe the formation of $^8$Be on the exit channel of the $\alpha$ transfer reaction. 
For this purpose, the repulsive core of the AB potential was slightly weakened 
to give the ($1s$) state $\Phi_\alpha(\bm{r}')$ of the $\alpha$ cluster in $^8$Be 
a quasi-binding energy of 0.01 MeV. The cluster wave functions $\Phi_{NL}(\bm{r})$ 
and $\Phi_\alpha(\bm{r}')$ are used explicitly in the calculation
of the complex nonlocal $\alpha$ transfer form factor   
\begin{equation}
\langle\beta'|W|\beta\rangle\sim \langle[\Phi_\alpha(\bm{r}')\otimes 
 Y_{L_{\beta'}}(\hat{{\bm R}'})]_{J_{\beta'}}|W|[\Phi_{NL}(\bm{r})\otimes 
 Y_{L_{\beta}}(\hat{\bm R})]_{J_{\beta}}\rangle, \label{eq13}
\end{equation}
where $L_{\beta}$ and $L_{\beta'}$ are the relative orbital momenta of the
initial and final partitions. The wave functions of $^8$Be core in the initial
and $^4$He core in the final partitions are omitted in (\ref{eq13}) because they 
are spectators and do not contribute to the transfer \cite{ThompsonN}. 

The CRC calculation of the $\alpha$ transfer $^{12}$C($\alpha$,$^8$Be) reaction 
requires the input of the spectroscopic factor of the $\alpha$ cluster in $^{8}$Be 
which is naturally assumed to be unity, and that of the cluster configuration 
$\alpha$+$^{8}$Be in $^{12}$C. The latter is determined as $S_\alpha=|A_{NL}|^2$, 
where the spectroscopic amplitude $A_{NL}$ is given by the dinuclear overlap  
\begin{equation}
\langle{\rm ^{8}Be}|^{12}{\rm C}\rangle=A_{NL}\Phi_{NL}(\bm{r}). \label{eq14}
\end{equation}     
Because two different exit channels of the $\alpha$ transfer $^{12}$C($\alpha$,$^8$Be) 
reaction were identified, with the emitting $^8$Be being in the g.s. and excited 
$2^+$ state \cite{Woz76}, one needs to evaluate (\ref{eq14}) for the two 
configurations $\alpha$+$^{8}$Be$_{\rm g.s.}$ and $\alpha$+$^{8}$Be$^*_{2^+}$, 
which are associated with $\Phi_{N=3,L=0}(\bm{r})$ ($S$-wave) and 
$\Phi_{N=2,L=2}(\bm{r})$ ($D$-wave). In general, one can treat these two $S_\alpha$ 
values as free parameters to be adjusted by the best DWBA or CRC fit to the $\alpha$
transfer data. Instead of this procedure, we have adopted in the present work the 
$S_\alpha$ values predicted for these configurations by Kurokawa and Kato using 
the CSM method \cite{KaKu07}. Namely, $S_\alpha ({\rm g.s.})\approx 0.36$ and 
$S_\alpha (2^+)\approx 0.38$, which are rather close to the spectroscopic factors 
predicted recently by other cluster models \cite{FHO09,Enyo07}. Note that the 
$S_\alpha$ values used in our CRC calculation are also close to those extracted 
from the DWBA analysis of the $^8$Be transfer reaction  
$^{24}$Mg($\alpha$,$^{12}$C)$^{16}$O \cite{BHJ80}. The same \BeBe OP has been used 
for both  $^8\text{Be}+^8$Be$_{\rm g.s.}$ and $^8\text{Be}+^8$Be$^*_{2^+}$ exit 
channels of the final partition (see more discussion below). 

\begin{figure}[bht]\vspace*{-1cm}
 \includegraphics[width=\textwidth]{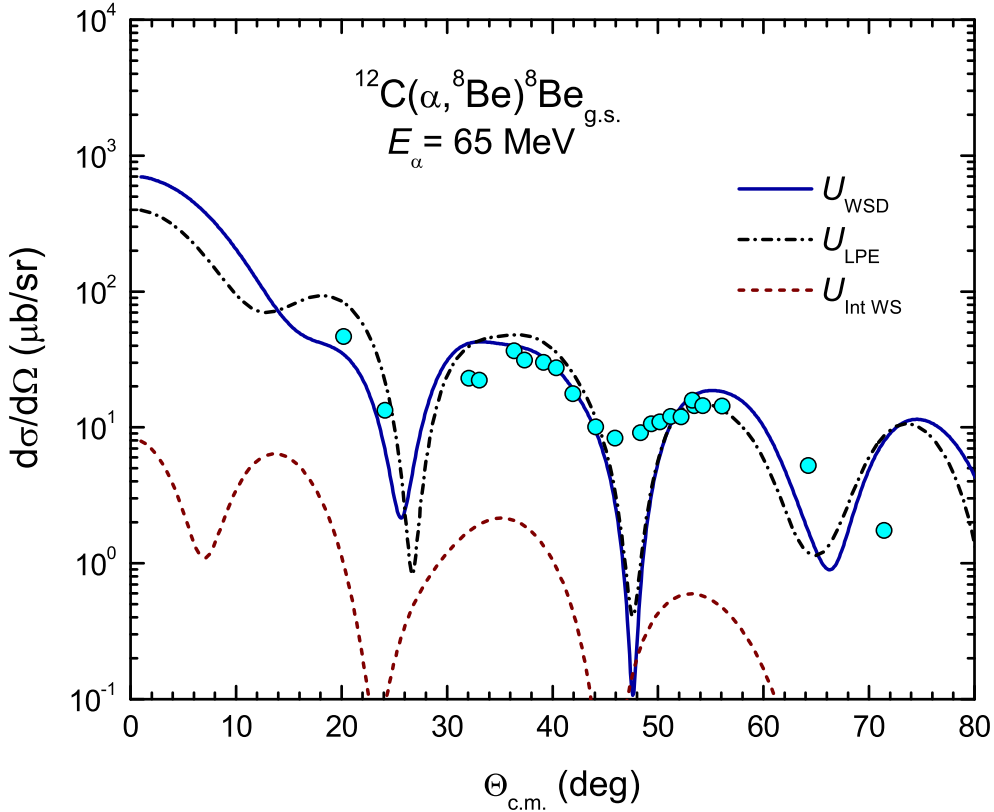}\vspace*{-0.5cm} 
 \caption{CRC description of the $\alpha$ transfer reaction 
$^{12}$C($\alpha$,$^{8}$Be)$^{8}$Be$_{\rm g.s.}$ measured at $E_{\alpha}=65$ MeV 
\cite{Woz76}, using the $\alpha$ spectroscopic factor $S_\alpha({\rm g.s.})\approx 0.36$ 
taken from the CSM calculation \cite{KaKu07}. The CRC results obtained with the CDCC-based
optical potentials $U_{\rm WSD}$ and $U_{\rm LEP}$ for the \BeBe partition are shown 
as the solid and dash-dotted lines, respectively. The dashed line is the CRC result 
obtained with the WS potential $U_{\rm Int WS}$, interpolated from the OP's adopted 
for the $^{7,8}$Be+$^9$Be systems at the nearby energies \cite{Rom09}.} \label{f6}
\end{figure}

The CRC results for the $\alpha$ transfer $^{12}$C($^4$He,$^8$Be) reaction at 
$E_{\alpha}=65$ MeV to the ground state of $^8$Be are compared with the measured data 
\cite{Woz76} in Fig.~\ref{f6}. One can see that the CDCC-based optical potentials
of the \BeBe partition ($U_{\rm LEP}$ and $U_{\rm WSD}$ determined by the methods 
(i) and (ii), respectively) give a good CRC description of the $\alpha$ transfer data 
without any adjustment of its strength, using 
$S_\alpha({\rm g.s.})=|A_{30}|^2\approx 0.36$ taken from the results of the CSM 
calculation \cite{KaKu07}. With a better OM description of the CDCC elastic cross 
section given by the $U_{\rm WSD}$ potential (see Fig.~\ref{f3}), the CRC cross section 
given by $U_{\rm WSD}$ also agrees slightly better the measured $\alpha$ transfer data.   

Because the \BeBe OP is unknown so far, a practical assumption is to estimate it 
from the phenomenological OP's adopted for the neighboring $^{9,7}$Be isotopes. 
For example, the proton transfer reaction $^7$Li($^{10}$B,$^9$Be)$^8$Be was measured 
by Romanyshyn {\it et al.} \cite{Rom09}, and a deep WS potential was deduced for 
the real OP of the $^8$Be+$^9$Be system at $E_{\rm c.m.}=31.7$ MeV from the DWBA 
analysis of transfer data. Interestingly, these authors also found that the 
$^8$Be+$^9$Be OP is quite close to that adopted earlier for the $^7$Be+$^9$Be 
system (see Fig.~10 of Ref.~\cite{Rom09}). Therefore, one might expect the \BeBe 
OP to be close to the WS optical potentials adopted for the $^{7,8}$Be+$^9$Be systems. 
To explore the reliability of this practical approach, we have interpolated
the \BeBe OP from those of the $^{7,8}$Be+$^9$Be systems adopted in Ref.~\cite{Rom09} 
and denoted it as $U_{\rm Int WS}$, with $V_v=155.0$ MeV, $R_v=3.152$ fm, $a_v=0.768$ fm; 
and $W_v=13.5$ MeV, $R_w=5.6$ fm, $a_w=0.768$ fm. The use of $U_{\rm Int WS}$ in the CRC 
calculation of the $^{12}$C($\alpha$,$^{8}$Be) reaction completely fails to account 
for the data (see dashed lines in Fig.~\ref{f6}). In fact, the spectroscopic factor 
$S_\alpha({\rm g.s.})$ taken from Ref.~\cite{KaKu07} must be scaled by a factor 
of 25, so that the CRC cross section obtained with $U_{\rm Int WS}$ can be comparable 
with the measured $\alpha$ transfer data. 
\begin{figure}[bht]\vspace*{-0.5cm}
 \includegraphics[width=\textwidth]{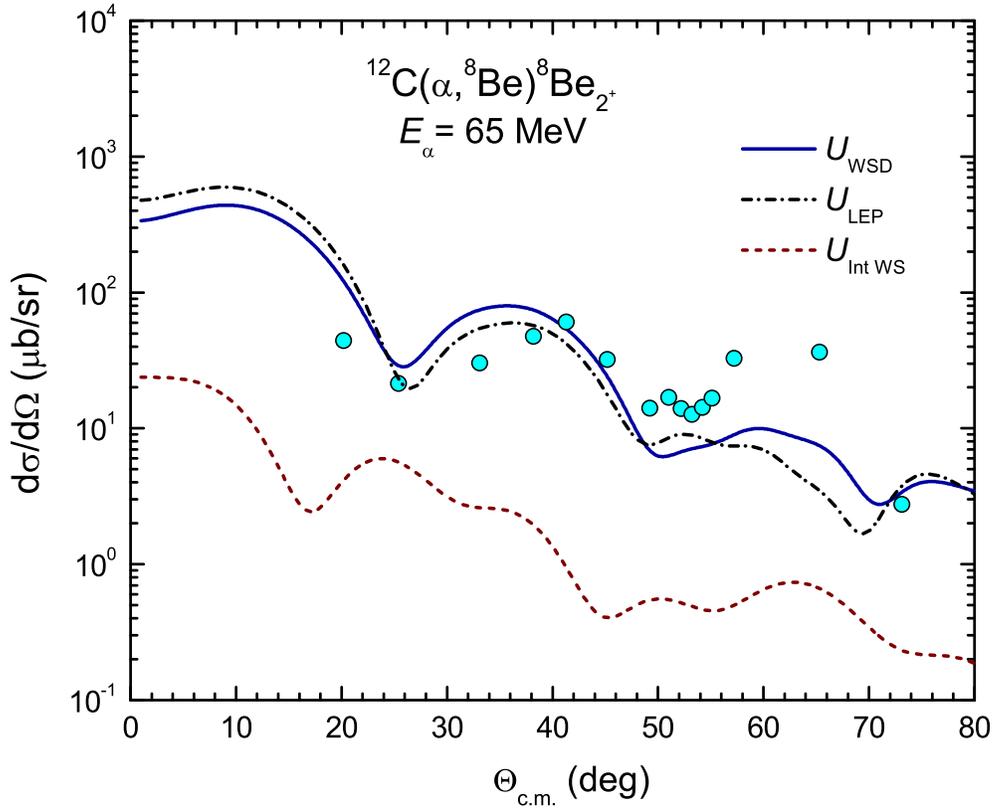}\vspace*{-0.5cm} 
 \caption{The same as Fig.~\ref{f6} but for the $\alpha$ transfer reaction with
 one emitting $^{8}$Be nucleus being in its $2^+$ state ($E_x\approx 2.94$ MeV),
 using the $\alpha$ spectroscopic factor $S_\alpha(2^+)\approx 0.38$ taken from 
 the CSM calculation \cite{KaKu07}} \label{f7}
\end{figure}

We have also performed the CRC calculation of the $^{12}$C($\alpha,^8$Be)$^8$Be$^*$ 
reaction at $E_{\alpha}=65$ MeV with one emitting $^8$Be nucleus being in its $2^+$ 
state ($E_x\approx 2.94$ MeV). Although the $2^+$ state of $^8$Be is a broad resonance, 
its 2$\alpha$-cluster structure remains similar to that of the ground state, and 
the $\alpha$ transfer cross section measured for the $2^+$ state is of about the same 
strength as that measured for the g.s. as shown in Figs.~\ref{f6} and \ref{f7}. 
The  $\alpha$ spectroscopic factors $S_\alpha$ predicted by the CSM calculation 
\cite{KaKu07} are also close for both the ground- and $2^+$ states.  It is, 
therefore, reasonable to use the same CDCC-based \BeBe OP for the partition 
$^8$Be+$^8$Be$_{2^+}^*$ in the exit channel. The results of the CRC calculation 
are compared with the data \cite{Woz76} in Fig.~\ref{f7}, and one can see that 
the (unrenormalized) CDCC-based \BeBe OP also delivers a good description 
of the $\alpha$ transfer data using $S_\alpha(2^+)=|A_{22}|^2\approx 0.38$ given 
by the CSM calculation \cite{KaKu07}. The use of $U_{\rm Int WS}$ for the \BeBe OP
in the CRC calculation also strongly underestimates the $\alpha$ transfer data 
(see dashed line in Fig.~\ref{f7}).

In conclusion, a good CRC description of the measured $^{12}$C($\alpha,^8$Be) data 
has been obtained with the \BeBe OP's determined by the methods (i) and (ii) from
the elastic \BeBe cross section given by the 4$\alpha$ CDCC calculation, and 
with the $\alpha$ spectroscopic factors given by the CSM calculation \cite{KaKu07}. 
The fact that no adjustment of the potential strength of $U_{\rm LEP}$ and 
$U_{\rm WSD}$ was necessary suggests that the 4$\alpha$ CDCC method is a reliable 
approach to study the \BeBe system. These results also show that, despite the short 
life-time of $^{8}$Be, the $\alpha$ transfer $^{12}$C($^4$He,$^8$Be) reaction is very 
sensitive to the OP of the \BeBe partition. The measured $\alpha$ transfer data 
clearly prefer the shallow OP based on the result of the 4$\alpha$ CDCC calculation 
using the AB potential of the \aa interaction \cite{AB66}, over the deep WS potential
interpolated from those adopted for the $^{7,8}$Be+$^9$Be systems \cite{Rom09}.         

\section{Summary}
The 3-body and recently suggested 4-body CDCC methods \cite{De18} have been used 
to predict the elastic \Bea and \BeBe scattering at $E_{\rm c.m.}=43.3$ and 41.3 
MeV, respectively, using the \aa interaction suggested by Ali and Bodmer \cite{AB66} 
that well reproduces the experimental \aa phase shifts. The elastic cross sections
predicted by the CDCC calculation were used to determine the local OP’s of the \Bea 
and \BeBe systems.  

The CDCC-based \Bea and \BeBe OP's are further used as the inputs of the core-core OP 
and that of the final partition, respectively, in the CRC study of the $\alpha$ 
transfer $^{12}$C($\alpha,^8$Be) reaction measured at $E_\alpha=65$ MeV \cite{Woz76}, 
with the emitting $^8$Be being in both the g.s. and $2^+$ state. These $\alpha$ transfer 
data are well reproduced by the CRC results obtained with the CDCC-based OP’s and 
$\alpha$ spectroscopic factors of the $^8$Be+$^8$Be$_{\rm g.s.}$ and
$^8$Be+$^8$Be$_{2^+}^*$ configurations in $^{12}$C taken from the CSM cluster 
calculation \cite{KaKu07}.

As alternative to the shallow (surface-type) \BeBe OP determined from the 
elastic \BeBe cross section predicted by the 4$\alpha$ CDCC calculation, a deep 
WS potential with parameters interpolated from the OP’s adopted for the 
$^{7,8}$Be+$^9$Be systems at the nearby energies \cite{Rom09} has been used 
in the CRC calculation, and it grossly underestimates the $\alpha$ transfer data. 
This might be due to the fact that $^7$Be and $^9$Be are well bound nuclei, and 
the breakup effect is therefore much weaker than that of the unbound $^8$Be. 

We conclude that the \Bea and \BeBe optical potentials can be determined from the
elastic scattering cross section predicted, respectively, by the 3$\alpha$ and 
4$\alpha$ CDCC calculations using the realistic \aa interaction that properly 
reproduces the experimental \aa phase shifts \cite{AB66}. The present CRC study should 
motivate further theoretical and experimental studies of the $\alpha$ transfer 
$^{12}$C($\alpha,^8$Be) reaction as a probe of the 4$\alpha$ interaction and 
the $\alpha$-cluster structure of $^{12}$C. 

\section*{Acknowledgement}
The present research has been supported, in part, by the National Foundation for
Scientific and Technological Development of Vietnam (NAFOSTED Project No. 103.04-2016.35). 
P.D. is Directeur de Recherches of F.R.S.-FNRS, Belgium, supported by the Fonds 
de la Recherche Scientifique - FNRS under Grant Number 4.45.10.08. Computational resources have
been provided by the Consortium des Équipements de Calcul
Intensif (CÉCI), funded by the Fonds de la Recherche Scientifique de Belgique (F.R.S.-FNRS) 
under Grant No. 2.5020.11 and by the Walloon Region. We also thank 
Nguyen T.T. Phuc for his helpful discussion. 

\bibliographystyle{apsrev4-1}
\bibliography{references_pk}
\end{document}